\documentclass[journal=jacsat,manuscript=article]{achemso}

\usepackage[version=3]{mhchem} 

\author{Chen Shen}
\affiliation{Institut f\"ur Materialwissenschaft, Technische Universit\"at Darmstadt, 64287, Darmstadt, Germany}

\author{Niloofar Hadaeghi}
\affiliation{Institut f\"ur Materialwissenschaft, Technische Universit\"at Darmstadt, 64287, Darmstadt, Germany}

\author{Harish K. Singh}
\affiliation{Institut f\"ur Materialwissenschaft, Technische Universit\"at Darmstadt, 64287, Darmstadt, Germany}

\author{Teng Long}
\affiliation{Institut f\"ur Materialwissenschaft, Technische Universit\"at Darmstadt, 64287, Darmstadt, Germany}

\author{Ling Fan}
\affiliation{Institute of Applied Materials, Karlsruhe Institute of Technology, 76131, Karlsruhe, Germany}

\author{Guangzhao Qin}
\affiliation{State Key Laboratory of Advanced Design and Manufacturing for Vehicle Body, College of Mechanical and Vehicle Engineering, Hunan University, 410082, Changsha, P. R. China}
\email{gzqin@hnu.edu.cn}

\author{Hongbin Zhang}
\affiliation{Institut f\"ur Materialwissenschaft, Technische Universit\"at Darmstadt, 64287, Darmstadt, Germany}
\email{hzhang@tmm.tu-darmstadt.de}

\title[An \textsf{achemso} demo]
  {Anomalously low thermal conductivity of two-dimensional GaP monolayers: A comparative study of the group GaX (X = N, P, As)}

\abbreviations{IR,NMR,UV}

\begin{document}

\begin{tocentry}

\centering 
\includegraphics[width=8cm, height=4cm]{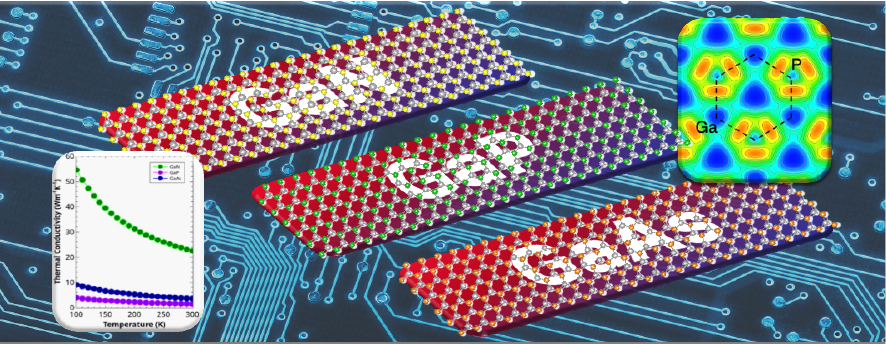}

\end{tocentry}

\begin{abstract}
With the successful synthesis of the two-dimensional (2D) gallium nitride (GaN) in a planar honeycomb structure, the phonon transport properties of 2D GaN have been reported. However, it remains unclear for the thermal transport in Ga-based materials by substituting N to other elements in the same main group, which is of more broad applications. In this paper, based on first-principles calculations, we performed a comprehensive study on the phonon transport properties of 2D GaX (X = N, P, and As) with planar or buckled honeycomb structures. The thermal conductivity of GaP (1.52 Wm$^{-1}$K$^{-1}$) is found unexpectedly ultra-low, which is in sharp contrast to GaN and GaAs despite their similar honeycomb geometry structure. Based on PJTE theory, GaP and GaAs stabilize in buckling structure, different from the planar structure of GaN. Compared to GaN and GaAs, strong phonon-phonon scattering is found in GaP due to the strongest phonon anharmonicity. Given electronic structures, deep insight is gained into the phonon transport that the delocalization of electrons in GaP is restricted due to the buckling structure. Thus, non-bonding lone pair electrons of P atoms induce nonlinear electrostatic forces upon thermal agitation, leading to increased phonon anharmonicity in the lattice, thus reducing thermal conductivity. Our study offers a fundamental understanding of phonon transport in GaX monolayers with honeycomb structure, which will enrich future studies of nanoscale phonon transport in 2D materials.
\end{abstract}

\section{Introduction}
The effective manipulation of thermal energy and thermal transport plays a pivotal role in the thermal management
for advanced energy and nano-electronic devices~\cite{balandin2011thermal,song2018two}.
On the one hand, materials with enhanced thermal transports are indispensable to maximize the heat transfer or minimize the heat waste, which can be applied to improve the working stability and energy efficiency of microelectronics.
On the other hand, systems with low thermal conductivity benefit the performance of the thermal barrier coating and thermoelectric devices~\cite{li2020recent}.
Therefore, insulators with tailored thermal transport properties originated from the crystal lattices are essential, as they can be integrated as thermal management components without causing other complications.~\cite{twaha2016comprehensive}
There is a strong impetus to gain deeper insights into the thermal transport mediated by phonons and to further treat the appealing thermophysical problems with enormous practical implications, which can be applied in electronic cooling~\cite{bistritzer2009electronic}, thermoelectrics~\cite{li2020recent}, phase change memories~\cite{balandin2012phononics,cahill2014nanoscale}, thermal devices (diodes, transistors, logic gates)~\cite{li2004thermal}, etc.

Particularly, initiated by the discovery of graphene~\cite{geim2009graphene},
2D materials have been intensively investigated for promising applications in 
engineering miniaturised devices.~\cite{balandin2012phononics,schedin2007detection}
For instance, a variety of 2D materials have been theoretically predicted and successfully fabricated, such as Xenes ({\it e.g.}, black phosphorus)~\cite{li2014black}, transition-metal dichalcogenides (TMDCs) ({\it e.g.}, MoS$_2$)~\cite{nie2017ultrafast}, MXenes (e.g., Ti$_3$C$_2$, and Ti$_4$N$_3$)~\cite{li2017mxene}, and nitrides ({\it e.g.}, BN)~\cite{watanabe2004direct}, 
which provide alternative solutions for electronic, spintronic, and optoelectronic applications. 
Moreover, the more efficient high-throughput density functional theory method is implemented to screen the novel 2D thermoelectric materials~\cite{sarikurt2020high}.
Recently, monolayer GaN with a planar honeycomb structure was successfully fabricated in experiments~\cite{al2016two,seo2015direct}and has been intensively theoretically studied~\cite{qin2017orbitally,qin2017anomalously}, which shows low thermal conductivity and is considered as potential application in energy conversion such as thermoelectrics. 
Therefore, an interesting question is how does the thermal transport perform in monolayer GaP and GaAs as the same main group of GaN and whether they are also good candidates as potential thermoelectric applications.

In this work, we performed first-principles calculations on the thermal conductivities in a series of Ga-based 2D materials GaX (X = N, P, and As).
It is observed that the lattice thermal conductivity of GaP monolayers is unexpectedly ultra-low, which is in sharp contrast to GaN and GaAs monolayers. 
Detailed analysis of the crystal structure and mode-resolved thermal conductivities reveals that the lone-pair non-bonding electrons play a critical role in the thermal conductivity. 
Such lone-pairs are strongly correlated with the crystal structure distortions, which can be attributed to the pseudo Jahn-Teller effect (PJTE).
Such mechanistic understanding of the thermal conductivities in GaX monolayers and the established electronic structure descriptors pave the way to optimize and design novel 2D materials as thermal functional materials and  to enrich the studies of nanoscale phonon transport in 2D materials.

\section{Computational details}
\textit{Ab initio} calculations based on density functional theory (DFT) were performed using the \textit{Vienna ab initio simulation package} (VASP)~\cite{kresse1993ab,kresse1996efficient}, which implements the projector augmented wave (PAW)~\cite{kresse1999ultrasoft}.
Exchange–correlation energy functional is treated using the Perdew–Burke–Ernzerhof of generalized gradient approximation (GGA-PBE)~\cite{perdew1996generalized}.
The wave functions are expanded in plane wave basis with a 20 $\times$ 20 $\times$ 1 Monkhorst-Pack~\cite{monkhorst1976special} k-sampling grid and cut-off energy of 1000 eV. 
A large vacuum region is set as 20 {\AA} to avoid the interactions between the monolayer and its mirrors induced by the periodic boundary conditions.
Precision of total energy convergence for the self-consistent field (SCF) calculations was as high as 10$^{-8}$ eV.
All structures are fully optimized until the  maximal Hellmann–Feynman force is less than 10$^{-8}$ eV{\AA}$^{-1}$.
To calculate the phonon dispersion, thermal conductivity, and various phonon properties, it is necessary to extract second- and third-interatomic force constants (IFCs) from first-principles calculations. 
To this end, 6 $\times$ 6 $\times$ 1 supercells containing 72 atoms were constructed, which is sufficiently large to allow the out-of-phase tilting motion.
To extract second-order IFCs, an atom in the supercell was displaced from its equilibrium position by 0.01 {\AA} and the Hellmann-Feynman forces were calculated based on the displaced configuration. 
Besides, the Born effective charges (Z$^*$) and dielectric constants ($\epsilon$) are obtained based on the density functional perturbation theory (DFPT), which is added to the dynamical matrix as a correction to take the long-range electrostatic interactions into account.
Lattice thermal conductivity ($\kappa_L$) and relative phonon properties were determined by solving the phonon BTE, as implemented in the ALAMODE~\cite{tadano2014anharmonic} package. 
The $\kappa_L$ is estimated by the BTE within RTA through the following equation:
\begin{equation}
    \kappa_L^{\alpha\beta}(T)=\frac{1}{NV}\sum_qC_q(T)\nu_q^{\alpha}(T)\nu_q^{\beta}(T)\tau_q(T),
\end{equation}
where $V$, $C_q(T)$, $\nu_q(T)$, and $\tau_q(T)$ are the unit cell volume, mode specific heat, phonon group velocity, and phonon lifetime, respectively.
When calculating the thermal conductivity, the real thickness of 2D materials can be obtained by considering the van der Waals radius of the upper and lower atoms plus the distance between the upper and lower atoms (buckling distance in this work).
At last, the stabilities of current systems can be checked in the C2DB database~\cite{haastrup2018computational}, which also indicates the GaSb and GaBi unstable.

\section{Results and discussion}
\subsection{Lattice structures of monolayer Ga-based compounds}
\begin{figure*}
\centering
  \includegraphics[width=15cm]{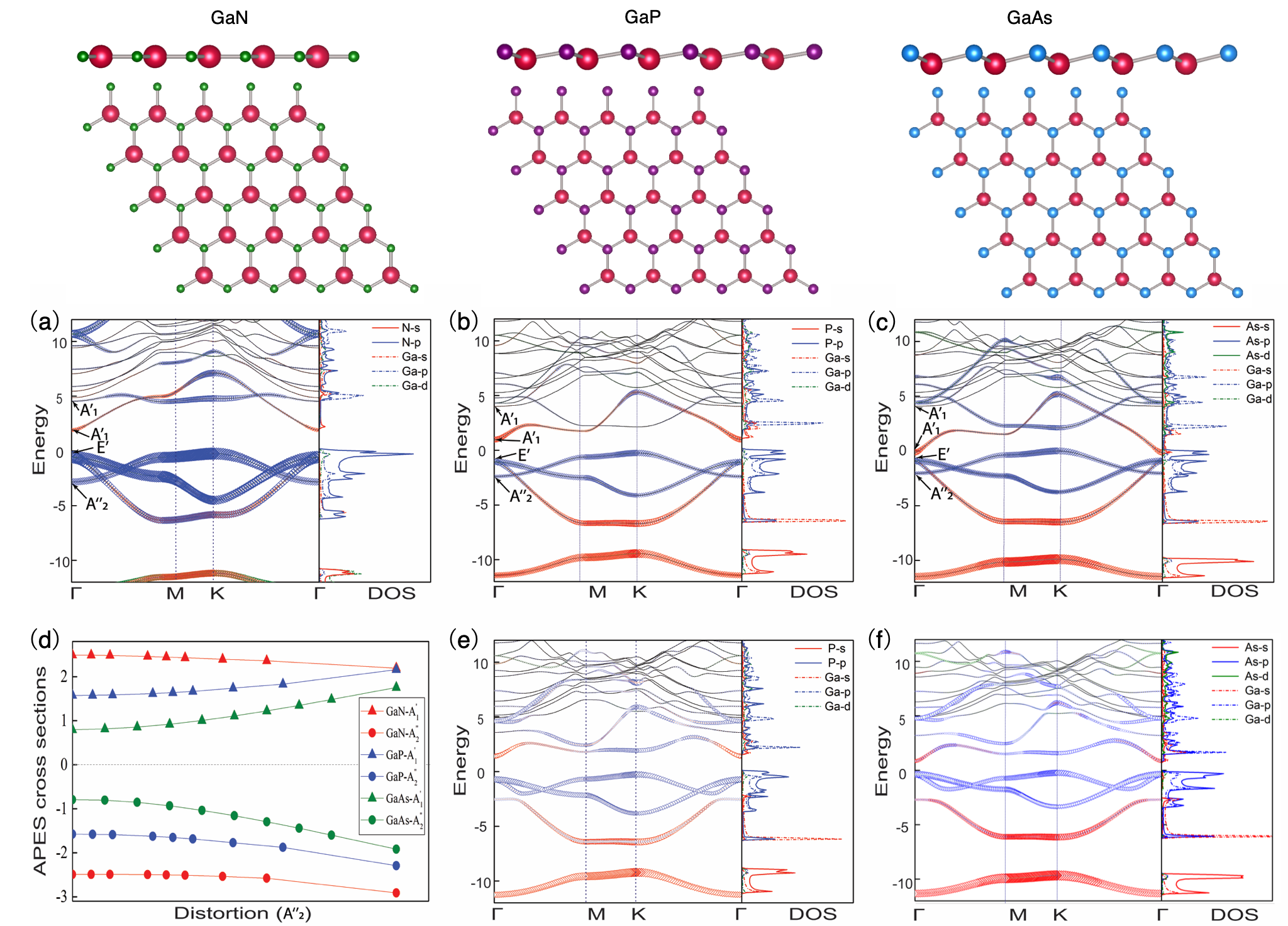}
  \caption{The side and top views of structures, phonon dispersions, and partial density of states (pDOS) of monolayer GaN, GaP and GaAs. Electronic band structures and pDOS of planar monolayer (a) GaN, (b) GaP, and (c) GaAs. (d) The adiabatic potential energy surface cross section of planar GaN, GaP, and GaAs with respect to the $A_2^{''}$ distortion mode. Electronic band structures and pDOS of buckled monolayer (e) GaP and (f) GaAs.}
  \label{fgr:eband}
\end{figure*}

The optimized structures of 2D GaN, GaP, and GaAs monolayers are shown in the Fig.~\ref{fgr:eband}. 
According to the top views, all three GaX monolayers exhibit honeycomb structures, with the resulting lattice parameters listed in Table~\ref{tbl:structure}. 
However, GaN monolayers have a planar structure, whereas the GaP and GaAs monolayers are buckled, {\it i.e.}, the Ga and X sublattices are shifted in opposite directions perpendicular to the monolayers, leading to larger thickness than GaN (as listed in Table~\ref{tbl:structure}).
The lattice constants of the GaN, GaP, and GaAs are in good agreement with previous reports~\cite{csahin2009monolayer,qin2017anomalously}.
For instance, the in-plane lattice constant of GaN is 3.21 {\AA} as optimized in this study, which is between 3.20 {\AA} in Ref.~\cite{csahin2009monolayer} and 3.26 {\AA} in Ref.~\cite{qin2017anomalously}.

\begin{table}[h!]
\small
  \caption{\ Symmetry space group, lattice constant (a in \AA), thickness (\AA), and buckling distance (\AA) of monolayer GaN, GaP, and GaAs. (The real thickness of 2D materials can be obtained by considering the van der Waals radius of the upper and lower atoms plus the distance between the upper and lower atoms.)}
  \label{tbl:structure}
  \begin{tabular*}{\textwidth}{@{\extracolsep{\fill}}llllc}
    \hline
    Compound & Space group & Thickness & a & Buckling distance\\
    \hline
    GaN & P$\bar{6}$m2 & 3.74 &3.21& 0\\
    GaP & P3m1 & 4.06 &3.90 & 0.39\\
    GaAs & P3m1 & 4.30 & 4.05&0.58\\
    \hline
  \end{tabular*}
\end{table} 

The origin of such differences ({\it i.e.}, planar versus buckled) in the crystal structures can be attributed to the electronic structure. 
Following the theory of PJTE, the curvature of the adiabatic potential energy surface (APES) yields
\begin{equation}
    K=\underbrace{\left \langle \Psi_0\left | (\frac{\partial ^2H}{\partial Q^2})_0 \right | \Psi_0\right \rangle}_{K_0} \underbrace{-2\sum_{n}\frac{\left | \left \langle \Psi_0 \left | (\frac{\partial H}{\partial Q})_0 \right |\Psi_n\right \rangle \right |^2}{E_n-E_0}}_{K_v}
\label{eq:pjte}
\end{equation}
where H represents the Hamiltonian, $\Psi_0$($\Psi_n$) denotes the ground state (excited state) wave function, and all the functions are considered for the high-symmetry configuration.
It is noted that for the high-symmetry configuration the $K_0$ term is greater than  zero,~\cite{opik1957studies,bersuker1984origin,bersuker1980activated} and 
$K_v$, the vibronic contribution, is smaller than zero and the source of instability.
If the resulting $K=K_0+K_v$ is smaller than zero, the crystal structure is unstable with respect to the distortion mode denoted by Q. 
Correspondingly, the allowed virtual transitions can be obtained based on the symmetry analysis, {\it i.e.}, the direct product of the irreducible representations (irrep) of the ground state $\Gamma_0$, the excited state $\Gamma_n$, and the distortion mode $\Gamma_q$ should contain the $A_{1g}$ representation. That is, $\Gamma_0\otimes \Gamma_q\otimes \Gamma_n \supset A_{1g}$.
In other words, when the direct product of the representations of the electronic states contains the irrep of the distortion mode,~\cite{powell2010symmetry} {\it i.e.}, $ \Gamma_0\otimes \Gamma_n \supset \Gamma_q$, there is possible finite contribution to the $K_v$ term. Besides, in the case of strong PJTE, the curvature of ground (excited) state in the APES becomes negative (positive) with respect to the q distortion mode.

For the GaX monolayers, following the PJTE at at the center of the BZ ($\Gamma$ point), the buckling is induced by the $A_2^{''}$ mode, which causes the phase transition from the high-symmetric P$\bar{6}$m2 structure to the low-symmetric P3m1 structure. Based on symmetry analysis and according to the irreps of the electronic states in Fig.~\ref{fgr:eband}(a-c), only the $A_2^{''}$ (originated mostly from the X-p states) and $A_1'$ (mainly of the X-s character) states are allowed to be coupled by the $A_2^{''}$ distortion mode, because $A_2'' \otimes A_1' = A_2''$.
The resulting adiabatic potential energy surface cross section with respect to the $A_2^{''}$ distortion amplitude is shown in Fig.~\ref{fgr:eband}(d). 
Obviously, the softening of the ground state, with $A_2^{''}$ irrep, and increasing of the excited state, with $A_1^\prime$ irrep, increases from the GaN to the GaAs monolayers.
One main reason is the reduced energy difference between the two electronic states ($E_n-E_0$ in the $K_V$ term), marked as energy gap, {\it e.g.},the energy gap changes from 4.98 eV for GaN, to 3.15 eV for GaP, and finally to 1.59 eV for GaAs monolayers.
Thus, such an enhanced PJT coupling leads to the buckling of the crystal lattices of GaP and GaAs.

\subsection{Phonon dispersion}
\begin{figure*}
 \centering
 \includegraphics[width=15cm]{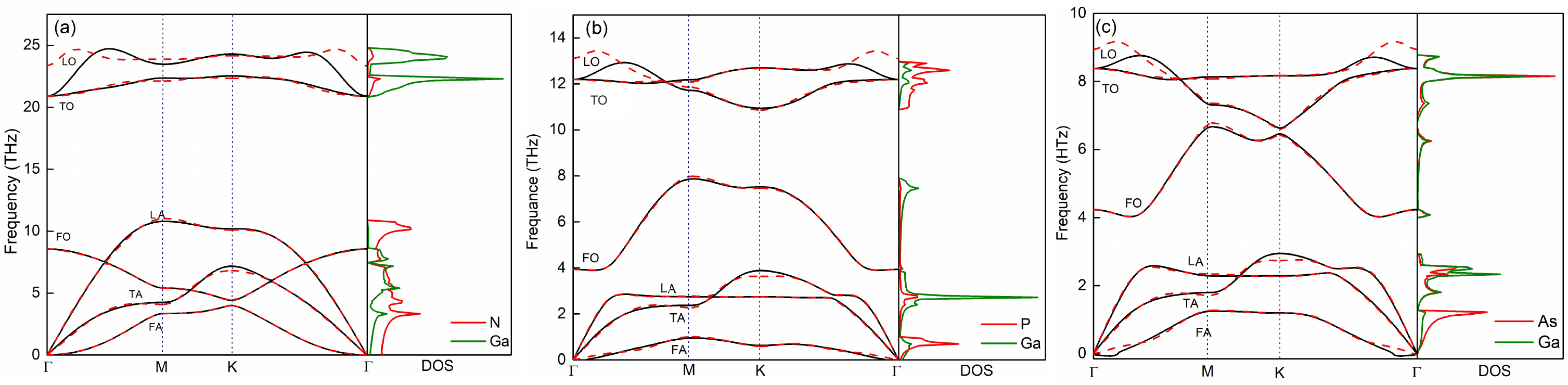}
 \caption{The phonon dispersion considering the effect of Born effective charges and dielectric constants is plotted in a violet dash-dot line, showing LO-TO splitting at the center of the BZ ($\Gamma$ point). The phonon dispersions not including the dipole correction are also plotted in solid line for comparison.}
 \label{fgr:structure}
\end{figure*}

Turning now to the lattice dynamics and thermal transport properties. 
Fig.~\ref{fgr:structure} shows the phonon spectra of GaX monolayers obtained obtained by diagonalizing the dynamical matrix based on the second order IFCs.
As the GaX systems behave more like ionic insulators, the longitudinal optical (LO) - transverse optical (TO) splitting clearly
occurs after considering the nonanalytical corrections based on the Born effective charges listed in Table~\ref{tbl:born}.
This indicates the arising of macroscopic electric fields resulted from the atomic displacements associated 
with the long-wave LO phonons.~\cite{gonze1997dynamical}
Note also that after considering the non-analytical correction, the slightly imaginary mode at {$\Gamma$} point for GaP and GaAs monolayers disappear. 
This is due to the macroscopic field generated by the strongly polarized covalent bonds, leading to modified force constants and hence dynamical stability at the $\Gamma$ point.

Interestingly, in comparison to those of GaP and GaAs monolayers,
there exists a big gap in the phonon spectra of GaN monolayers between the LO/TO and other phonon bands.
This can be attributed to the large difference in the atomic mass of Ga and N atoms.  
Moreover, the frequencies of the acoustic branches of GaP and GaAs monolayers are lower than those in GaN cases, 
and much lower than the common 2D materials, such as \textit{h}-BN~\cite{qin2018lone} and graphene~\cite{lindsay2010flexural}.
This implies that the phonon harmonic vibrations of GaP and GaAs are weak, which will have a significant effect on the phonon transport properties.
In addition, the longitudinal acoustic (LA) and transverse acoustic (TA) phonon branches of the GaX systems present linear behavior when approaching to the $\Gamma$ point, while the flexural acoustic (FA) phonon branch shows a quadratic behavior. This is consistent with our previous results in GaN~\cite{qin2017anomalously}, which is a common behavior for 2D materials.

\begin{table*}
\small
  \caption{\ Born effective charges (Z$^*$) of Ga and X (where X = N, P, and As) atoms and the
dielectric constants ($\epsilon$) of GaP, GaP, and GaAs.}
  \label{tbl:born}
  \begin{tabular*}{0.95\textwidth}{@{\extracolsep{\fill}}llll}
    \hline
    GaN & Z$^*$(Ga) & Z$^*$(N) & $\epsilon$\\
    \hline
    xx & 3.071 & -3.071 & 1.859 \\
    yy & 3.071 & -3.071 & 1.859 \\
    zz & 0.337 & -0.337 & 1.148 \\
    \hline
    GaP & Z$^*$(Ga) & Z$^*$(P)  & $\epsilon$\\
    \hline
    xx &  2.975 & -2.975 & 3.025 \\
    yy &  2.975 & -2.975 & 3.025 \\
    zz &  0.156 & -0.156 & 1.179 \\
    \hline

    GaAs  & Z$^*$(Ga)& Z$^*$(As)& $\epsilon$\\
    \hline
    xx & 2.955&-2.955&3.901\\
    yy & 2.955&-2.955&3.901\\
    zz & 0.107 &-0.107&1.193\\
    \hline
  \end{tabular*}
\end{table*}

\subsection{Anomalous thermal conductivity}
\begin{figure*}[h]
\centering
  \includegraphics[width=15cm]{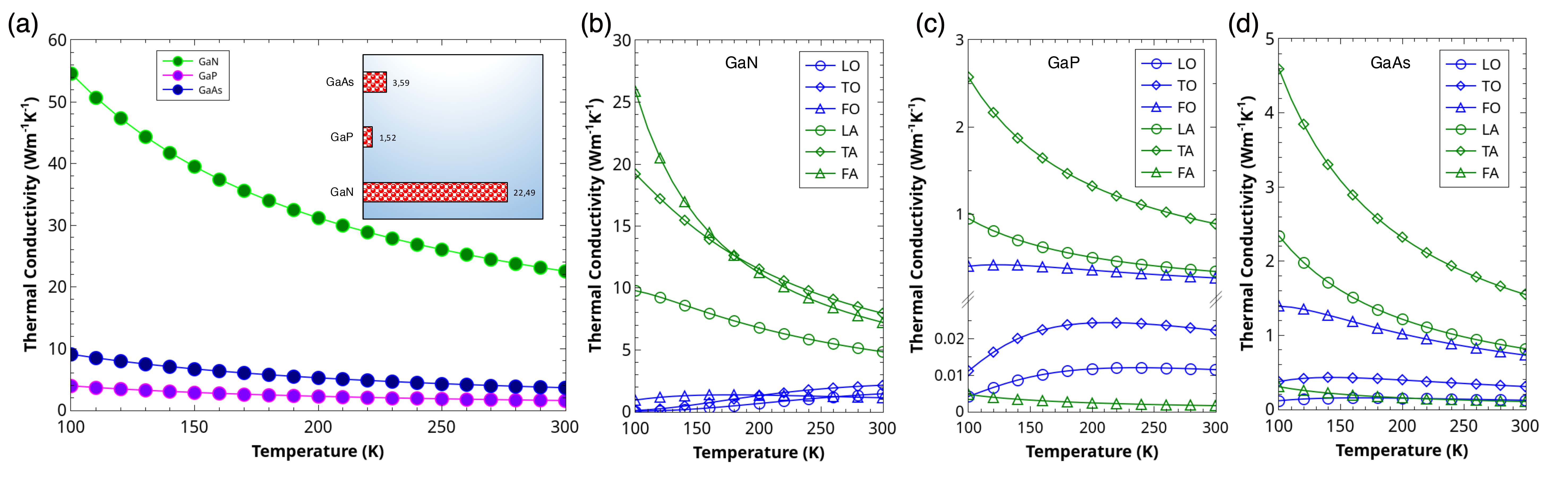}
  \caption{(a) Temperature (100-300 K) dependent thermal conductivities of monolayer GaN, GaP, and GaAs. Inset figure at the right corner shows the thermal conductivity (Wm$^{-1}$K$^{-1}$) of the three compounds at 300 K. (b-d) The absolute contribution to the total conductivity of monolayer GaN, GaP, and GaAs from each individual phonon branch as a function of temperature}
  \label{fgr:TC}
\end{figure*}

Fig.~\ref{fgr:TC} (a) shows the thermal conductivities of the three systems as a function of temperature, evaluated by solving the BTE with Born effective charges considered.
Clearly, the temperature dependence of the lattice thermal conductivities presents the typical 1/T behavior, consistent with other crystalline materials in both bulk and 2D forms.
Furthermore, GaN has the highest thermal conductivities in the whole temperature range, while those of GaP and GaAs are on average more than five times smaller.
The most striking result illustrated in Fig.~\ref{fgr:TC} (a) is that the thermal conductivity shows a non-monotonous behaviour when moving from N to As, {\it i.e.}, GaP monolayers possesses the lowest thermal conductivity. 
For instance, at 300 K, the thermal conductivity of GaP monolayers is 1.52 Wm$^{-1}$K$^{-1}$, 
which is half of the value of GaAs monolayers and more than one order of magnitude smaller than that of GaN monolayers.
Similarly, Sun {\it et al.}~\cite{sun2020ultra} reported ultra-low thermal conductivities for 2D triphosphides (InP$_3$, GaP$_3$, SbP$_3$, and SnP$_3$), which might be driven by the flatter acoustic phonon branches as expected for the GaP and GaAs monolayers.
It is noted that the thermal conductivities of GaP and GaAs monolayers are even lower than those of typical 2D thermoelectric materials such as GeSe~\cite{hao2016computational} and SnSe~\cite{zhao2016snse}.

To understand the underlying mechanism responsible for the ultra-low thermal conductivities of GaP monolayers and the anomalous trend for the GaX series, the mode-resolved  (FA, TA, LA, FO, TO, and LO modes) contributions to the thermal conductivity are shown in the Fig.~\ref{fgr:TC} (b-d) for GaX monolayers.
Obviously, the acoustic modes exhibit dominant contributions in contrast to the optical modes.
Moreover, the FA and TA branches make the most significant contributions to the total thermal conductivity of GaN monolayers, while the TA branch dominates the phonon transport in GaP and GaAs monolayers. 
The reason for the domination of the FA branch in GaN has been analyzed in previous work,~\cite{qin2017anomalously} 
where the reflectional symmetry of the planar honeycomb structure of GaN monolayers leads to the symmetry-based selection rule of phonon-phonon scattering and results in the small scattering rate of FA phonons~\cite{lindsay2010flexural}. 
On the contrary, the FA branch has drastically reduced contribution to the total thermal conductivities of GaP and GaAs monolayers due to the buckled (non-planar) crystal structures.

\subsection{Mode level analysis}
\begin{figure}[h]
\centering
  \includegraphics[width=15cm]{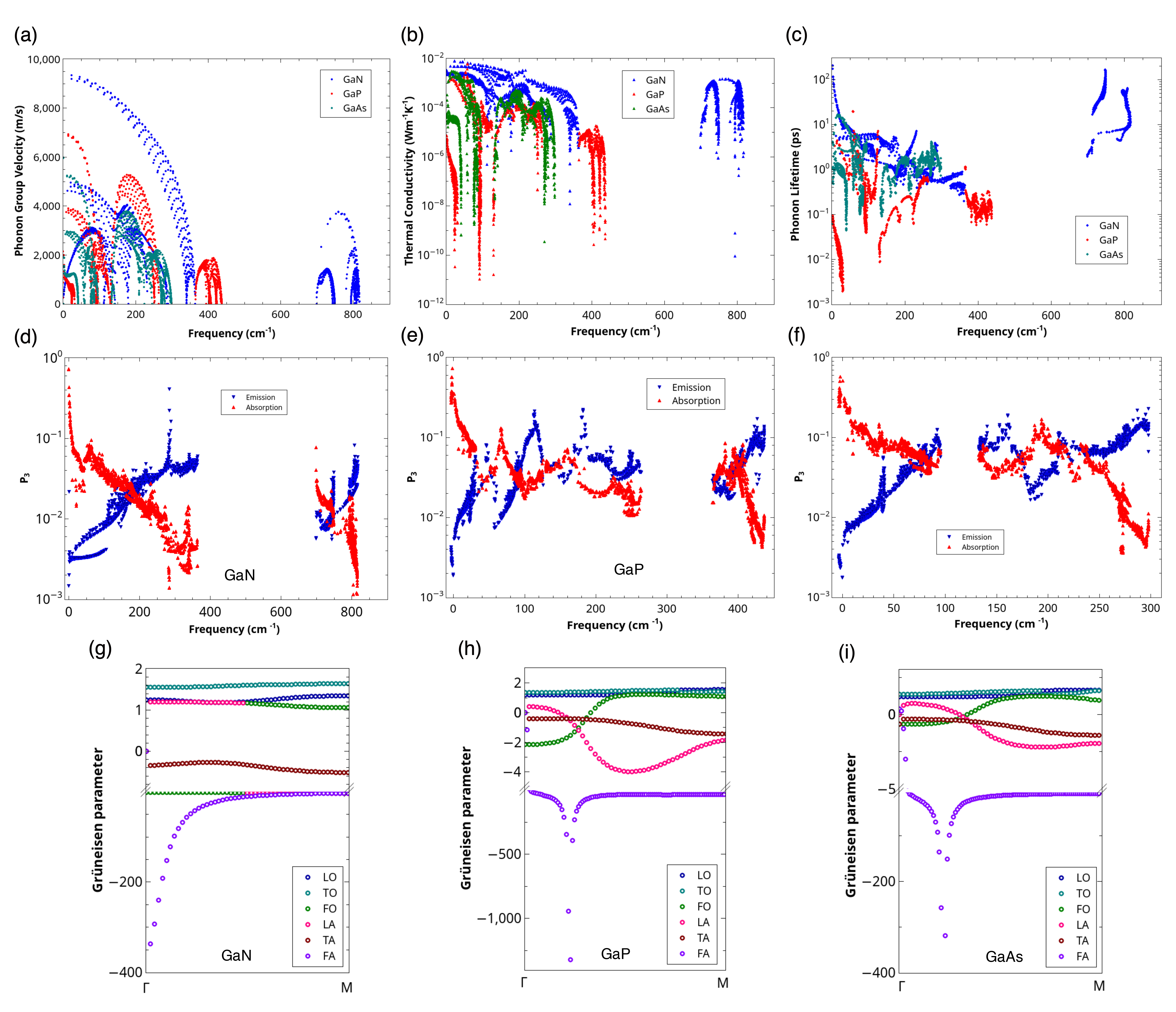}
  \caption{(a) The comparison of mode phonon group velocity, the comparison of the mode-level (b) contributions to thermal conductivity and (c) phonon lifetime of monolayer GaN, GaP, and GaAs at 300K. (d-f) The mode-level scattering phase space of absorption and emission processes, and (g-i) the mode level Grüneisen parameters for three compounds.}
  \label{fgr:mode-level}
\end{figure}

\begin{figure}[h]
\centering
  \includegraphics[width=15cm]{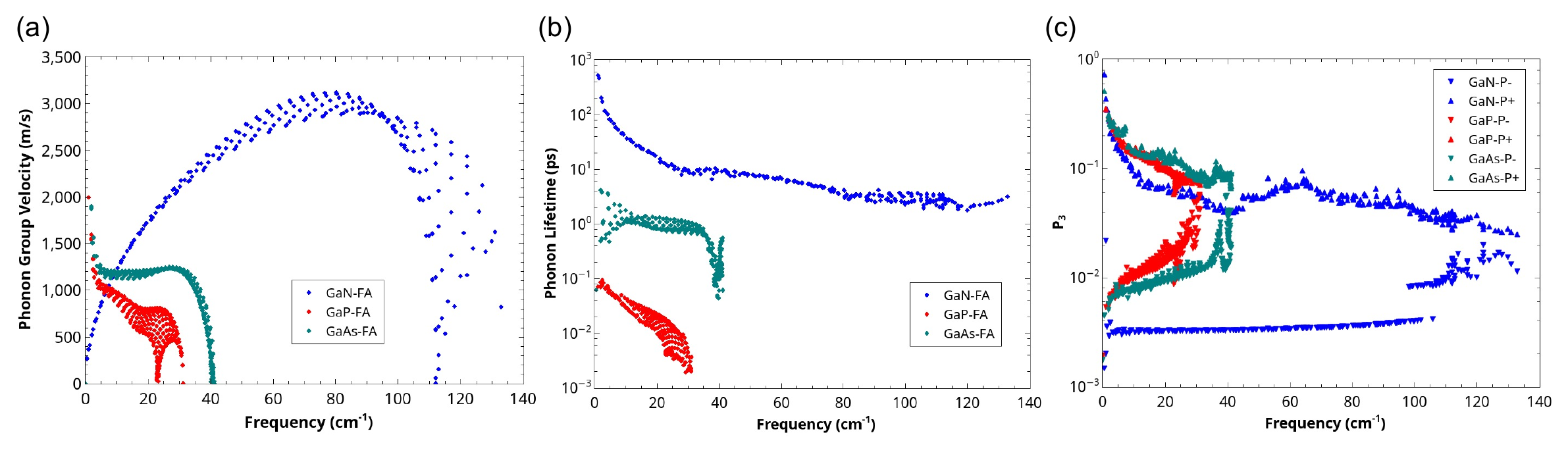}
  \caption{The comparison of (a) phonon group velocity, (b) phonon lifetime, and (c) the scattering phase space of absorption (P+) and emission (P-) processes of FA mode of GaX monolayers.}
  \label{fgr:FA}
\end{figure}

To gain further insight into the thermal transport in GaX monolayers, we performed detailed analysis on the mode level phonon properties.
Comparison of the mode level phonon group velocity of GaN, GaP and GaAs as a function of frequency at 300 K are shown in Fig.~\ref{fgr:mode-level} (a). 
It is clearly seen that the overall phonon group velocity of monolayer GaP and GaAs are on the same order of magnitude, which is smaller than that of monolayer GaN. 
Note that, the optical phonon branches of GaN and GaAs have relatively larger group velocities. 
Besides, the phonon velocity of FO branches of GaN, GaP and GaAs are large among the other branches
Interestingly, the contribution of thermal conductivity of FA mode for GaX monolayers are dramatically different, as mentioned in previous section. Hence, we plotted the phonon propeties of FA branch in the Fig.~\ref{fgr:FA}. Note that, the phonon group velocity of FA mode of GaN has larger values and wilder distribution than those of GaP and GaAs. Especially, the phonon group velocity of FA mode of GaP is concentrated in a smaller value area, leading to the lowest thermal conductivity.

The mode-level contribution to thermal conductivity and the corresponding phonon lifetime at 300 K are shown in the Fig.~\ref{fgr:mode-level} (b,c).
It can be found that the phonon frequencies of monolayer GaN contributing to the thermal conductivity are concentrated at 0-400 cm$^{-1}$, which is consistent with the results as illustrated in Fig.~\ref{fgr:TC}. 
And the main contribution of phonon frequencies in GaP and GaN are distributed at a much lower range, which can be confirmed by the phonon dispersion as shown in Fig.~\ref{fgr:structure}.
It is worth noting that not only acoustic branches, but also some optical branches with low-frequency play a major role in contributing to the thermal conductivity, which is more pronounced in GaN.
Commonly, the phonons with low frequency dominating the main contribution of thermal conductivity is the universal for 2D materials.
Fig.~\ref{fgr:mode-level} (c) shows the phonon lifetime for GaN, GaP and GaAs at 300 K. 
Compared to GaN and GaAs, GaP has the lowest phonon relaxation time, which could be an indicator of the strong phonon anharmonicity, leading to an ultra-low lattice thermal conductivity.
For GaAs, the phonon relaxation time in the high frequency range above the gap is comparable with that in the low frequency range below the gap.
For GaN, the phonon lifetime of optical phonon branches are quite high, some of them are even larger than acoustic phonon modes. Thus, the thermal conductivity of GaN is much higher than the other two.
As same as the group velocity of FA mode for GaX monolayers, the FA mode in GaP has the lowest phonon lifetime (as shown in the Fig.~\ref{fgr:FA} (b)), which also results in the lowest thermal conductivity of GaP in these three materials.

Furthermore, based on phonon dispersion, the scattering phase space has been calculated with the criteria of energy and momentum conservation. 
As shown in Fig.~\ref{fgr:mode-level}, the mode level scattering phase space of
GaN, GaP and GaAs are presented for the phonon modes available for absorption
and emission processes, respectively.
It is clearly seen that there is an inverse relationship between phase space for the three-phonon process.
As one can see, because of the selection rule applied to the planar honeycomb structure of monolayer GaN, the phonon scattering processes involving odd numbers of FA phonon modes are largely restricted, which results in the dominant contribution of the FA mode to thermal conductivity. 
For the buckled GaP and GaAs where symmetry-based selection rule of phonon-phonon scattering are broken, the scattering phase space of FA branches available for both absorption and emission processes is not reduced consequently.
As shown in Fig.~Fig.~\ref{fgr:FA} (c), the scattering phase space for absorption processes of FA modes of GaP and GaAs are larger, which subsequently leads to the less contribution to thermal conductivity.

As one known, the phonon-phonon scattering process is influenced by the anharmonic nature of structures, which can be roughly quantified by the Grüneisen parameter.
In this vein, we analyzed the phonon anharmonicity of the GaN, GaP and GaAs by calculating the Grüneisen parameter.
As shown in the Fig.~\ref{fgr:mode-level} (g-i), the magnitude of Grüneisen parameter of GaP is much larger than that of other two, especially for the FA phonon branch, which indicates the strongest phonon anharmonicity in GaP. 
Owing to the strong anharmonicity, the strong phonon-phonon scattering results in the small phonon lifetime, as previously shown in Fig. \ref{fgr:mode-level} (c).
Hence, monolayer GaP has the ultra-low thermal conductivity.

\subsection{Insight from electronic structures}
The systematic investigation of model level phonon transport in the framework of Boltzmann transport theory has been implemented in the above sections to analyze the ultra-low thermal conductivity of monolayer GaP due to strong phonon anharmonicity.
In this section, we conduct intensive study on the electronic structures to get deep insight into the phonon transport and the phonon anharmonicity.
We will present that the active lone-pair electrons due to special orbital hybridization and buckling structures drive the remarkable phonon anharmonicity in monolayer GaP.

\begin{figure}
\centering
  \includegraphics[width=15cm]{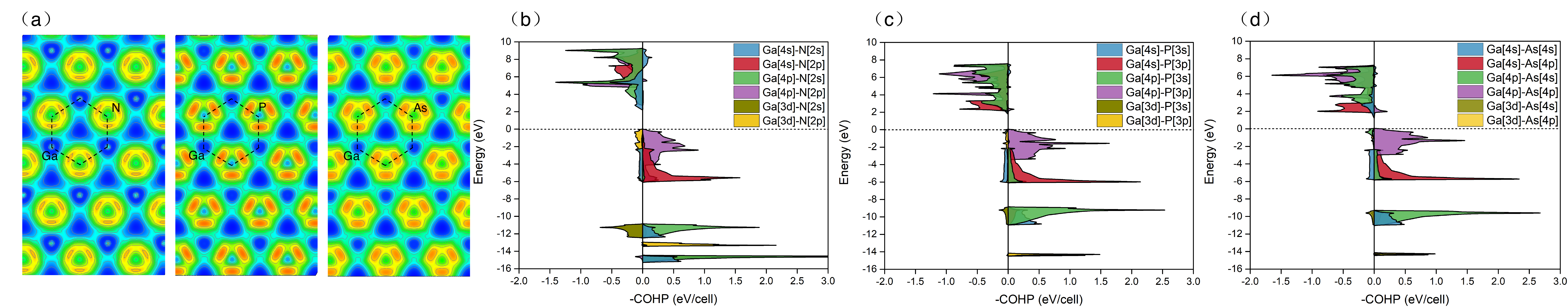}
  \caption{ (a) The top view of electron localization functions (ELF), and (b-d) Orbital-resolved COHP of monolayer GaN, GaP, and GaAs}
  \label{fgr:elf}
\end{figure}

It was proposed by Petrov and Shtrum that lone-pair electrons could lead to low thermal conductivity~\cite{petrov1962heat}.
As shown in the Fig.~\ref{fgr:elf} (a), the electron localization functions (ELF) provides information on the structure of atomic shells, and also displays the location and size of bonding and lone electron pairs~\cite{savin1997elf}.
Non-bonding lone-pair electrons arise around N, P, and As atoms.
The overlapping wave functions of lone-pair electrons with valence electrons from adjacent atoms Ga induce nonlinear electrostatic forces upon thermal agitation, which leads to increased phonon anharmonicity in the lattice and thus reduces the thermal conductivity~\cite{petrov1962heat,xiao2016origin,zeier2016thinking, jana2016origin,nielsen2013lone,skoug2011role,morelli2008intrinsically}.

It is noted that there exists difference in the electronegativity between Ga and (N, P, and As), which leads to polarization for the Ga-(N, P, and As) bonds as revealed by ELF. 
Considering the largest difference in the electronegativity between Ga and N atoms, the phonon anharmonicity in GaN is expected to be the strongest. 
However, this does not happen because GaP and GaAs have relatively lower thermal conductivities.
The reason might be that the buckled honeycomb structures of GaP and GaAs would induce more ionicity and restrict the delocalization of electrons, leading to the much stronger localization of lone-pairs electrons as shown in the Fig.~\ref{fgr:elf} (a).
Moreover, the bonding electrons for the Ga-P (As) bonds is relatively closer to P (As) atom, which contributes positively to the stronger interaction with the non-bonding P(As)-\textit{s} electrons and thus leads to a stronger phonon anharmonicity.

To learn more about the bonding formability with respect to the variation of orbital states of atoms, detailed analysis on crystal orbital Hamilton population (COHP) is carried out.
A positive -pCOHP value indicate the bonding interaction, while a negative value indicates the antibonding interaction.
Thus, the active lone-pair electrons are considered as neither bonding nor anti-bonding interactions.
The integrated COHP of orbitals should be zero closing to Fermi energy.
Generally, the active lone-pair electrons are dominantly contributed by the \textit{s}-orbital. 
In this regard, we focus on the \textit{4s} orbital of Ga and outermost \textit{s} orbital of N, P, and As.
As shown in Fig.~\ref{fgr:elf} (b-d), closing to the Fermi level, Ga[p]-(N, P, and As)[p] and Ga[s]-(N, P, and As)[p] orbitals hybridize and dominantly contribute to the bonding, which indicates a positive -pCOHP at Fermi level.
In contrast, the (N, P, and As)[s]-Ga[p,d] orbitals present neither bonding nor anti-bonding interactions.
Therefore, N, P, and As form the polar covalent bonds with Ga by sharing the $p$ electrons, while the $s^2$ electrons of N, P, and As form isolated (lone pair) bands.
Such behavior is also confirmed by the electronic band structures and partial density of states as shown in the Fig.~\ref{fgr:eband}. 
The $s$ orbital is largely (around 10~eV) confined below the valence band, forming an isolated band. 
However, the situation for the orbitals is different for the N atom where the \textit{s} orbitals hybrid with Ga-\textit{d} orbitals. 
In this regards, we can find relatively large anti-bonding Ga[3d]-N[2s] as shown in the Fig.~\ref{fgr:elf} (b).

In this vein, due to the buckling structures, the delocalization of electrons in
GaP and GaAs are restricted. The non-bonding lone pair electron of P and As
atoms are more stronger, which induce nonlinear electrostatic forces upon
thermal agitation, leading to increased phonon anharmonicity in the lattice and
thus reducing the thermal conductivity.


\section{Conclusions}
In summary, by solving the phonon BTE based on first-principles calculations, we have performed a comprehensive study on the phonon transport properties of 2D GaX with planar and buckled honeycomb structures. The thermal conductivity of GaP is calculated to be 1.52 Wm$^{-1}$K$^{-1}$, which is unexpectedly ultra-low and in sharp contrast to GaN and GaAs. 
Considering the similar honeycomb geometry structure of GaP to that of GaN and GaAs, it is quite intriguing to find that the thermal conductivity of GaP is very low.
Firstly, to understand the underlying mechanism for GaX monolayers having planar or buckling structures, systematic analysis is performed based on PJTE theory.
The larger bandgap and smaller the vibronic coupling constant,  the less destabilization of the ground state and less stabilization of the excited states. Hence, the GaN exists in planar structure, and GaP and GaAs stabilize in buckling structures.
Then, in order to gain insight into anomalous phenomena of ultra-low thermal conductivity for GaP, we perform a detailed analysis of the underlying mechanisms in the framework of phonon mode-solved investigation.
The root reason for the low thermal conductivity the of GaP is found to be that: FA dominates the thermal conductivity of GaN but less contributes to the one of GaP, which is due to the symmetry-based selection rule and difference of atomic structure. In particular, the difference originates from the different situations for the phonon lifetime, which is determined by phonon–phonon scattering.
The phonon anharmonicity quantified by the Grüneisen parameter is further analyzed to understand the phonon–phonon scattering, indicating the strong phonon-phonon scattering of GaP and the strongest phonon anharmonicity of GaP in GaX. 
Considering that all the properties are fundamentally determined by the atomic structure and the behavior of electrons (such as charge distribution and orbital hybridization), we further perform analysis from the view of electronic structures and orbital bonding to gain deep insight into the phonon transport. Due to the buckling structures, the delocalization of electrons in GaP and GaAs are restricted. The non-bonding lone pair electron of P and As atoms are more stronger, which induce nonlinear electrostatic forces upon thermal agitation, leading to increased phonon anharmonicity in the lattice and thus reducing the thermal conductivity.
Our study offers fundamental understanding of phonon transport in GaX monolayers with honeycomb structure within the framework of BTE and the electronic structure from the bottom, which will enrich the studies of nanoscale phonon transport in 2D materials.
\section*{Conflicts of interest}
There are no conflicts to declare.

\begin{acknowledgement}

This work was supported by the Deutsche Forschungsgemeinschaft (DFG, German Research Foundation) - Project-ID 405553726 - TRR 270.
G.Q. is supported by the Fundamental Research Funds for the Central Universities (Grant No. 531118010471) and the National Natural Science Foundation of China (Grant No. 52006057).
The Lichtenberg high performance computer of the TU Darmstadt is gratefully acknowledged for the computational resources where the calculations were conducted for this project.

\end{acknowledgement}


\bibliography{achemso-demo}

\begin{thebibliography}{10}

\bibitem{balandin2011thermal}
A.~A. Balandin, ``Thermal properties of graphene and nanostructured carbon
  materials,'' {\em Nat. Mater.}, vol.~10, no.~8, pp.~569--581, 2011.

\bibitem{song2018two}
H.~Song, J.~Liu, B.~Liu, J.~Wu, H.-M. Cheng, and F.~Kang, ``Two-dimensional
  materials for thermal management applications,'' {\em Joule}, vol.~2, no.~3,
  pp.~442--463, 2018.

\bibitem{li2020recent}
D.~Li, Y.~Gong, Y.~Chen, J.~Lin, Q.~Khan, Y.~Zhang, Y.~Li, H.~Zhang, and
  H.~Xie, ``Recent progress of two-dimensional thermoelectric materials,'' {\em
  Nanomicro Lett.}, vol.~12, no.~1, pp.~1--40, 2020.

\bibitem{twaha2016comprehensive}
S.~Twaha, J.~Zhu, Y.~Yan, and B.~Li, ``A comprehensive review of thermoelectric
  technology: Materials, applications, modelling and performance improvement,''
  {\em Renew. Sustain. Energy Rev.}, vol.~65, pp.~698--726, 2016.

\bibitem{bistritzer2009electronic}
R.~Bistritzer and A.~H. MacDonald, ``Electronic cooling in graphene,'' {\em
  Phys. Rev. Lett}, vol.~102, no.~20, p.~206410, 2009.

\bibitem{balandin2012phononics}
A.~A. Balandin and D.~L. Nika, ``Phononics in low-dimensional materials,'' {\em
  Mater. Today}, vol.~15, no.~6, pp.~266--275, 2012.

\bibitem{cahill2014nanoscale}
D.~G. Cahill, P.~V. Braun, G.~Chen, D.~R. Clarke, S.~Fan, K.~E. Goodson,
  P.~Keblinski, W.~P. King, G.~D. Mahan, A.~Majumdar, {\em et~al.}, ``Nanoscale
  thermal transport. {II}. 2003--2012,'' {\em Appl. Phys. Rev.}, vol.~1, no.~1,
  p.~011305, 2014.

\bibitem{li2004thermal}
B.~Li, L.~Wang, and G.~Casati, ``Thermal diode: Rectification of heat flux,''
  {\em Phys. Rev. Lett}, vol.~93, no.~18, p.~184301, 2004.

\bibitem{geim2009graphene}
A.~K. Geim, ``Graphene: status and prospects,'' {\em Science}, vol.~324,
  no.~5934, pp.~1530--1534, 2009.

\bibitem{schedin2007detection}
F.~Schedin, A.~K. Geim, S.~V. Morozov, E.~Hill, P.~Blake, M.~Katsnelson, and
  K.~S. Novoselov, ``Detection of individual gas molecules adsorbed on
  graphene,'' {\em Nat. Mater.}, vol.~6, no.~9, pp.~652--655, 2007.

\bibitem{li2014black}
L.~Li, Y.~Yu, G.~J. Ye, Q.~Ge, X.~Ou, H.~Wu, D.~Feng, X.~H. Chen, and Y.~Zhang,
  ``Black phosphorus field-effect transistors,'' {\em Nat. Nanotechnol.},
  vol.~9, no.~5, p.~372, 2014.

\bibitem{nie2017ultrafast}
C.~Nie, L.~Yu, X.~Wei, J.~Shen, W.~Lu, W.~Chen, S.~Feng, and H.~Shi,
  ``Ultrafast growth of large-area monolayer {M}o{S}$_2$ film via gold foil
  assistant {CVD} for a highly sensitive photodetector,'' {\em Nanotechnology},
  vol.~28, no.~27, p.~275203, 2017.

\bibitem{li2017mxene}
R.~Li, L.~Zhang, L.~Shi, and P.~Wang, ``Mxene {T}i$_3${C}$_2$: an effective
  2{D} light-to-heat conversion material,'' {\em ACS nano}, vol.~11, no.~4,
  pp.~3752--3759, 2017.

\bibitem{watanabe2004direct}
K.~Watanabe, T.~Taniguchi, and H.~Kanda, ``Direct-bandgap properties and
  evidence for ultraviolet lasing of hexagonal boron nitride single crystal,''
  {\em Nat. Mater.}, vol.~3, no.~6, pp.~404--409, 2004.

\bibitem{sarikurt2020high}
S.~Sarikurt, T.~Kocaba{\c{s}}, and C.~Sevik, ``High-throughput computational
  screening of 2{D} materials for thermoelectrics,'' {\em J. Mater. Chem. A},
  vol.~8, no.~37, pp.~19674--19683, 2020.

\bibitem{al2016two}
Z.~Y. Al~Balushi, K.~Wang, R.~K. Ghosh, R.~A. Vil{\'a}, S.~M. Eichfeld, J.~D.
  Caldwell, X.~Qin, Y.-C. Lin, P.~A. DeSario, G.~Stone, {\em et~al.},
  ``Two-dimensional gallium nitride realized via graphene encapsulation,'' {\em
  Nat. Mater.}, vol.~15, no.~11, pp.~1166--1171, 2016.

\bibitem{seo2015direct}
T.~H. Seo, A.~H. Park, S.~Park, Y.~H. Kim, G.~H. Lee, M.~J. Kim, M.~S. Jeong,
  Y.~H. Lee, Y.-B. Hahn, and E.-K. Suh, ``Direct growth of gan layer on carbon
  nanotube-graphene hybrid structure and its application for light emitting
  diodes,'' {\em Sci. Rep.}, vol.~5, p.~7747, 2015.

\bibitem{qin2017orbitally}
Z.~Qin, G.~Qin, X.~Zuo, Z.~Xiong, and M.~Hu, ``Orbitally driven low thermal
  conductivity of monolayer gallium nitride ({G}a{N}) with planar honeycomb
  structure: a comparative study,'' {\em Nanoscale}, vol.~9, no.~12,
  pp.~4295--4309, 2017.

\bibitem{qin2017anomalously}
G.~Qin, Z.~Qin, H.~Wang, and M.~Hu, ``Anomalously temperature-dependent thermal
  conductivity of monolayer {G}a{N} with large deviations from the traditional
  1/{T} law,'' {\em Phys. Rev. B}, vol.~95, no.~19, p.~195416, 2017.

\bibitem{perdew1996generalized}
J.~P. Perdew, K.~Burke, and M.~Ernzerhof, ``Generalized gradient approximation
  made simple,'' {\em Phys. Rev. Lett}, vol.~77, no.~18, p.~3865, 1996.

\bibitem{haastrup2018computational}
S.~Haastrup, M.~Strange, M.~Pandey, T.~Deilmann, P.~S. Schmidt, N.~F. Hinsche,
  M.~N. Gjerding, D.~Torelli, P.~M. Larsen, A.~C. Riis-Jensen, {\em et~al.},
  ``The computational 2{D} materials database: high-throughput modeling and
  discovery of atomically thin crystals,'' {\em 2D Mater.}, vol.~5, no.~4,
  p.~042002, 2018.

\bibitem{csahin2009monolayer}
H.~{\c{S}}ahin, S.~Cahangirov, M.~Topsakal, E.~Bekaroglu, E.~Akturk, R.~T.
  Senger, and S.~Ciraci, ``Monolayer honeycomb structures of group-{IV}
  elements and {III-V} binary compounds: {F}irst-principles calculations,''
  {\em Phys. Rev. B}, vol.~80, no.~15, p.~155453, 2009.

\bibitem{gonze1997dynamical}
X.~Gonze and C.~Lee, ``Dynamical matrices, {B}orn effective charges, dielectric
  permittivity tensors, and interatomic force constants from density-functional
  perturbation theory,'' {\em Phys. Rev. B}, vol.~55, no.~16, p.~10355, 1997.

\bibitem{sun2020ultra}
Z.~Sun, K.~Yuan, Z.~Chang, S.~Bi, X.~Zhang, and D.~Tang, ``Ultra-low thermal
  conductivity and high thermoelectric performance of two-dimensional
  triphosphides ({I}n{P}$_3$, {G}a{P}$_3$, {S}b{P}$_3$ and {S}n{P}$_3$): {A}
  comprehensive first-principles study,'' {\em Nanoscale}, 2020.

\bibitem{lindsay2010flexural}
L.~Lindsay, D.~Broido, and N.~Mingo, ``Flexural phonons and thermal transport
  in graphene,'' {\em Phys. Rev. B}, vol.~82, no.~11, p.~115427, 2010.

\bibitem{opik1957studies}
U.~{\"O}pik and M.~H.~L. Pryce, ``Studies of the {J}ahn-{T}eller effect. {I}.
  {A} survey of the static problem,'' {\em Proceedings of the Royal Society of
  London. Series A. Mathematical and Physical Sciences}, vol.~238, no.~1215,
  pp.~425--447, 1957.

\bibitem{bersuker1984origin}
I.~B. Bersuker, N.~N. Gorinchoi, and V.~Z. Polinger, ``On the origin of dynamic
  instability of molecular systems,'' {\em Theor. Chim. Acta}, vol.~66,
  no.~3-4, pp.~161--172, 1984.

\bibitem{bersuker1980activated}
I.~Bersuker and B.~IB, ``Are activated complexes of chemical reactions
  experimentally observable ones.,'' 1980.

\bibitem{powell2010symmetry}
R.~C. Powell, {\em Symmetry, group theory, and the physical properties of
  crystals}, vol.~172.
\newblock Springer, 2010.

\bibitem{savin1997elf}
A.~Savin, R.~Nesper, S.~Wengert, and T.~F. F{\"a}ssler, ``{ELF}: The electron
  localization function,'' {\em Angew. Chem}, vol.~36, no.~17, pp.~1808--1832,
  1997.

\bibitem{kresse1996efficient}
G.~Kresse and J.~Furthm{\"u}ller, ``Efficient iterative schemes for ab initio
  total-energy calculations using a plane-wave basis set,'' {\em Phys. Rev. B},
  vol.~54, no.~16, p.~11169, 1996.

\bibitem{kresse1993ab}
G.~Kresse and J.~Hafner, ``Ab initio molecular dynamics for liquid metals,''
  {\em Phys. Rev. B}, vol.~47, no.~1, p.~558, 1993.

\bibitem{kresse1999ultrasoft}
G.~Kresse and D.~Joubert, ``From ultrasoft pseudopotentials to the projector
  augmented-wave method,'' {\em Phys. Rev. B}, vol.~59, no.~3, p.~1758, 1999.

\bibitem{tadano2014anharmonic}
T.~Tadano, Y.~Gohda, and S.~Tsuneyuki, ``Anharmonic force constants extracted
  from first-principles molecular dynamics: applications to heat transfer
  simulations,'' {\em J. Condens. Matter Phys.}, vol.~26, no.~22, p.~225402,
  2014.

\bibitem{monkhorst1976special}
H.~J. Monkhorst and J.~D. Pack, ``Special points for {B}rillouin-zone
  integrations,'' {\em Phys. Rev. B}, vol.~13, no.~12, p.~5188, 1976.

\bibitem{qin2018lone}
G.~Qin, Z.~Qin, H.~Wang, and M.~Hu, ``Lone-pair electrons induced anomalous
  enhancement of thermal transport in strained planar two-dimensional
  materials,'' {\em Nano Energy}, vol.~50, pp.~425--430, 2018.

\bibitem{hao2016computational}
S.~Hao, F.~Shi, V.~P. Dravid, M.~G. Kanatzidis, and C.~Wolverton,
  ``Computational prediction of high thermoelectric performance in hole doped
  layered gese,'' {\em Chem. Mater.}, vol.~28, no.~9, pp.~3218--3226, 2016.

\bibitem{zhao2016snse}
L.-D. Zhao, C.~Chang, G.~Tan, and M.~G. Kanatzidis, ``{S}n{S}e: a remarkable
  new thermoelectric material,'' {\em Energy Environ. Sci.}, vol.~9, no.~10,
  pp.~3044--3060, 2016.

\bibitem{petrov1962heat}
A.~Petrov and E.~Shtrum, ``Heat conductivity and the chemical bond in
  {ABX}$_2$-type compounds,'' {\em Soviet Physics-Solid state}, vol.~4, no.~6,
  pp.~1061--1065, 1962.

\bibitem{morelli2008intrinsically}
D.~Morelli, V.~Jovovic, and J.~Heremans, ``Intrinsically minimal thermal
  conductivity in cubic {I-V-VI} semiconductors,'' {\em Phys. Rev. Lett},
  vol.~101, no.~3, p.~035901, 2008.

\bibitem{skoug2011role}
E.~J. Skoug and D.~T. Morelli, ``Role of lone-pair electrons in producing
  minimum thermal conductivity in nitrogen-group chalcogenide compounds,'' {\em
  Phys. Rev. Lett}, vol.~107, no.~23, p.~235901, 2011.

\bibitem{nielsen2013lone}
M.~D. Nielsen, V.~Ozolins, and J.~P. Heremans, ``Lone pair electrons minimize
  lattice thermal conductivity,'' {\em Energy Environ. Sci.}, vol.~6, no.~2,
  pp.~570--578, 2013.

\bibitem{jana2016origin}
M.~K. Jana, K.~Pal, U.~V. Waghmare, and K.~Biswas, ``The origin of ultralow
  thermal conductivity in {I}n{T}e: {L}one-pair-induced anharmonic rattling,''
  {\em Angew. Chem}, vol.~128, no.~27, pp.~7923--7927, 2016.

\bibitem{xiao2016origin}
Y.~Xiao, C.~Chang, Y.~Pei, D.~Wu, K.~Peng, X.~Zhou, S.~Gong, J.~He, Y.~Zhang,
  Z.~Zeng, {\em et~al.}, ``Origin of low thermal conductivity in {S}n{S}e,''
  {\em Phys. Rev. B}, vol.~94, no.~12, p.~125203, 2016.

\bibitem{zeier2016thinking}
W.~G. Zeier, A.~Zevalkink, Z.~M. Gibbs, G.~Hautier, M.~G. Kanatzidis, and G.~J.
  Snyder, ``Thinking like a chemist: intuition in thermoelectric materials,''
  {\em Angew. Chem}, vol.~55, no.~24, pp.~6826--6841, 2016.

\end{thebibliography}

\end{document}